\documentclass[aps,floatfix,prstab,preprint,a4paper,%
superscriptaddress,showpacs]{revtex4}
\usepackage{amsfonts}
\usepackage[T1]{fontenc}
\usepackage{textcomp}
\usepackage[varg]{txfonts}
\newcommand\REM[1]{}
\usepackage{color}

\begin{document}

\title{Once more on the equivalence between quantum phase transition phenomena in
radiation-matter and magnetic systems}
\author{J. G. Brankov}
\email{brankov@bas.bg} \affiliation{Institute of Mechanics, acad G
Bonchev 4, 1113 Sofia, Bulgaria}

\author{N.S. Tonchev}
\email{tonchev@issp.bas.bg} \affiliation{Institute of Solid State
Physics, 72 Tzarigradsko Chauss\'ee, 1784 Sofia, Bulgaria}

\author{V. A. Zagrebnov}
\email{zagrebnov@cpt.univ-mrs.fr}
 \affiliation{Universit\'e de la
Mediterran\'ee (Aix-Marseille II) and Centre de Physique
Th\'eorique, Luminy-Case 907, 13288 Marseille, Cedex 09, France}

\begin{abstract}

In answer to the replies of Reslen {\it et al} [arXiv:
quant-ph/0507164 (2005)], and Liberti and Zaffino
[arXiv:cond-mat/0507019, (2005)], we comment once more on the
temperature-dependent effective Hamiltonians for the Dicke model
derived by them in [Europhys. Lett., {\bf 69} (2005) 8] and [Eur.
Phys. J., {\bf 44} (2005) 535], respectively. These approximate
Hamiltonians cannot be correct for any finite nonzero temperature
because they both violate a rigorous result. The fact that the
Dicke model belongs to the universality class of, and its
thermodynamics is described by the infinitely coordinated
transverse-field XY model is known for more than 30 years.
\end{abstract}


\maketitle

Quite recently and unexpectedly, a discussion on the equivalence
of the Dicke model Hamiltonian and effective spin-spin exchange
Hamiltonians emerged \cite{R05}-\cite{LZ}. As it is well known,
the Dicke model which describes the interaction of a single mode
radiation field with a system of $N$ two-level atoms (qubits) is
given by the Hamiltonian ($\hbar=c=\omega=1)$:
\begin{equation}\label{D}
H_{Dicke}=a^{\dagger}a + \epsilon J_{z} -
\left[\frac{2\lambda}{N^{1/2}}\right](a^{\dagger} + a)J_{x},
\end{equation}
where $J_{z} = \frac{1}{2}\sum_{i=1}^{N}\sigma_{i,z}$, $J_{x} =
\frac{1}{2}\sum_{i=1}^{N}(\sigma^{\dagger}_i +\sigma_i)$.

Instead of Eq. (\ref{D}), in Ref. \cite{R05} the following explicitly
temperature-dependent Hamiltonian is suggested as a lowest-order
approximation, which is claimed to provide "an ideal
starting point for studying the quantum phase transition (QFT) of the
Dicke model":
\begin{equation}\label{CT}
H_{qb}^{2}(\beta)=\epsilon J_{z} -
\left[\frac{2\lambda}{N^{1/2}}\right]^{2}J_{x}^{2}-
\left[\frac{2\lambda}{N^{1/2}}\right]^{2}
\frac{2}{\beta(h(\beta)+1)}J_{x}^{2}.
\end{equation}
Here $h(\beta)=(e^{\beta}-1)^{-1}$ is the Bose factor which determines
the average photon number in an isolated cavity (single radiation mode
of energy $\omega =1$) at inverse temperature $\beta =(k_{\rm B}T)^{-1}$.

A little bit later, in Ref. \cite{LZ05} another "high-temperature"
approximation is suggested for studying thermodynamic properties
of model (\ref{D}), which actually differs from Eq.(\ref{CT}) by
the replacement of the last term in that equation by the following
two terms:
\begin{equation}
\label{LZ}
+ \left[\frac{2\lambda}{N^{1/2}}\right]^{2}J_{x}^{2}-
\frac{\beta}{2}\left[\frac{2\lambda}{N^{1/2}}\right]^{2}\coth\left(
\frac{\beta}{2}\right)J_{x}^{2}.
\end{equation}
This temperature-dependent effective Hamiltonian is claimed to be "correct
for $\beta^3 \lambda^2 < 1$ and $\beta \epsilon \ll 1$, i.e. only in the
high-temperature limit".

In our comment \cite{BTZ} we have noted that, as a matter of fact, the above
effective Hamiltonians are {\it wrong for any finite nonzero temperatures}.
Because, {\it it is known for already more than 30 years that not only the
universality class of the Dicke model, but also the
thermodynamics of the effective qubit (spin) subsystem (including the
critical temperature, critical exponents, etc.) are described by the
thermodynamically equivalent Hamiltonian}
\begin{equation}\label{AH}
\left. H_{s}\right|_{\mu=1}=\epsilon J_{z} -
\left[\frac{2\lambda}{N^{1/2}}\right]^{2}J_{x}^{2}.
\end{equation}
This fact \textit{completely invalidates} the assertions of Ref.
\cite{R05} that:  "An immediate and interesting consequence is that
the exchange interaction can be controlled experimentally simply by
changing the system's temperature or, equivalently, the mean number
of photons in the cavity", and  "Temperature dependent phenomena can
also be studied using our size-consistent, effective qubit-qubit
system approach". This also means that the following pretension of
Ref. \cite{R05}: "We have also shown that the Dicke model pertains
to the same universality class as other infinitely-coordinated
systems, such as the anisotropic XY model in a transverse field." is
\textit{wrong = not true}.

First of all,
because it is \textit{not the authors} of \cite{R05} who have shown
this, and the second, because they never studied the Dicke model,
but a restricted \textit{caricature} on it.

Once more we emphasize that our statement has been proved rigorously
to hold exactly in the thermodynamic limit. Moreover, (i) its
exactness \textit{does not depend} on the values of the model
parameters $\epsilon$ and $\lambda$ and (ii) can be used in the
\textit{whole temperature interval} $0 \leq T < \infty$. In Ref.
\cite{BTZ} we have given an incomplete list of other publications,
supporting our statement by using different methods. At wish that
list can be made much longer. It is a matter of common sense, at
least, to use the \textit{correct} and, in addition, much more
simple effective Hamiltonian (\ref{AH}) instead of the
\textit{misguiding} approximations derived in Refs. \cite{R05} and
\cite{LZ05}.

In conclusion, our comment is very relevant to the core results of
the Refs. \cite{R05} and \cite{LZ05}, since, if their authors were
able to perform explicitly their perturbation procedures to any
order, they would inevitably arrive at Hamiltonian (\ref{AH}) -
because of its exactness, provided the corresponding expansion
series were convergent.

\end{document}